\documentclass[twocolumn,prl,showpacs]{revtex4}
\usepackage{graphicx}
\usepackage{dcolumn}
\usepackage{bm}
\usepackage{color}

\begin{document}
\title{Phase diagram of the one-dimensional half-filled extended Hubbard model}
\author{Satoshi Ejima$^1$ and Satoshi Nishimoto$^2$}
\affiliation{$^1$Fachbereich Physik, Philipps-Universit\"at Marburg, D-35032 Marburg, Germany \\
$^2$Max-Planck-Institut f\"ur Physik komplexer Systeme, D-01187 Dresden, Germany}
\date{\today}
\begin{abstract}

We study the ground state of the one-dimensional half-filled Hubbard model with on-site 
(nearest-neighbor) repulsive interaction $U$ ($V$) and nearest-neighbor hopping $t$. 
In order to obtain an accurate phase diagram, we consider various physical quantities such 
as the charge gap, spin gap, Luttinger-liquid exponents, and bond-order-wave (BOW) order 
parameter using the density-matrix renormalization group technique. We confirm that the 
BOW phase appears in a substantial region between the charge-density-wave (CDW) and 
spin-density-wave phases. Each phase boundary is determined by multiple means and it allows 
us to do a cross-check to demonstrate the validity of our estimations. Thus, our results 
agree quantitatively with the renormalization group results in the weak-coupling regime 
($U \lesssim 2t$), with the perturbation results in the strong-coupling regime ($U \gtrsim 6t$), 
and with the quantum Monte Carlo results in the intermediate-coupling regime. We also find 
that the BOW-CDW transition changes from continuous to first order at the tricritical point 
$(U_{\rm t}, V_{\rm t}) \approx (5.89t, 3.10t)$ and the BOW phase vanishes at the critical 
end point $(U_{\rm c}, V_{\rm c}) \approx (9.25t, 4.76t)$.
\end{abstract}
\maketitle

For several decades quasi-one-dimensional (1D) materials, e.g., organic 
conductors~\cite{ishiguro90}, conjugated polymers~\cite{kiess92}, and carbon 
nanotubes~\cite{ishii03}, have been a main subject of research in the field of condensed 
matter physics. A minimal electronic model which can describe their basic properties is 
the 1D extended Hubbard model (EHM)~\cite{baeriswyl85}. The Hamiltonian is given by
\begin{eqnarray}
\nonumber
H = &-&t \sum_{i,\sigma}(c^\dagger_{i \sigma}c_{i+1 \sigma} + H.c.) \\
&+&U \sum_i n_{i \uparrow}n_{i \downarrow} + V \sum_{i \sigma \sigma^\prime} n_{i \sigma}n_{i+1 \sigma^\prime},
\label{hamiltonian}
\end{eqnarray}
where $c_{i \sigma}^\dagger$ ($c_{i \sigma}$) is creation (annihilation) operator of 
an electron with spin $\sigma$ at site $i$, and $n_{i\sigma}=c_{i \sigma}^\dagger c_{i \sigma}$ 
is number operator. $t$ is nearest-neighbor hopping term and $U$ ($V$) is on-site 
(nearest-neighbor) Coulomb interaction. Despite the geometric simplicity, this model at half filling 
is believed to exhibit a variety of phases due to strong quantum fluctuations. 

Within the g-ology scheme~\cite{emery79}, the system has merely two insulating phases when 
the interaction strengths are positive: for $U<2V$ the ground state is 
$2k_{\rm F}$-charge-density-wave (CDW), where both the charge and spin excitations are gapped; 
for $U>2V$ a Mott insulator with $2k_{\rm F}$-spin-density-wave (SDW), where the spin 
excitation has no gap. However, based on non-perturbative numerical results, Nakamura argued 
that there is also a bond-order-wave (BOW) phase, where the ground state has a long-range 
staggered bond order, between the CDW and SDW phases~\cite{nakamura99}. 
So far much effort has been devoted to fix the ground-state phase diagram both 
analytically~\cite{hirsch84,cannon90,vandongen94,Voit95,tsuchiizu02,Tam06} and 
numerically~\cite{eric02,sengupta02,sandvik04,zhang04,Glocke07}. Nevertheless, surprisingly 
their results are in few (quantitative) agreements with each other. The aim of this paper 
is to produce  a highly accurate phase diagram of the 1D half-filled EHM and to resolve 
the apparent contradictions.

We employ the density-matrix renormalization group (DMRG) method, which is one of the most powerful numerical techniques for studying 1D many-body systems~\cite{white92}. With open-end 
boundary conditions, ground-state and low-lying excited-states energies as well as expectation 
values of physical quantities can be obtained quite accurately for very large finite-size systems 
(up to sites $L \sim {\cal O}(1000)$). In DMRG procedure we keep $m=1200$ to $3000$ 
density-matrix eigenstates, which are much larger than those in the previous DMRG 
studies~\cite{eric02,zhang04,Glocke07}, and all the calculated quantities are extrapolated to 
the $m \to \infty$ limit. In this way, the maximum  truncation  error, i.e., the  discarded 
weight, is less than $1 \times 10^{-11}$, while the maximum error in the ground-state energy 
is $\Delta E/t \sim 10^{-8}-10^{-7}$. We strongly argue that such large $m$ values and the 
$m$-extrapolation are essential for required accuracy of the measurements.

In order to determine the phase diagram including two phase boundaries, we calculate several 
physical quantities. Each boundary is determined by multiple means from the quantities 
and it allows us to do a cross-check on the estimates. First, to obtain the BOW-CDW 
boundary we calculate the charge gap
\begin{equation}
\Delta_{\rm c}=\lim_{L \to \infty}[E(N+2,0)+E(N-2,0)-2E(N,0)]/2,
\end{equation}
where $E(N_{\rm e},S_{\rm z})$ is the ground-state energy for a given number of electrons 
$N_{\rm e}$ and z-component of total spin $S_{\rm z}$. We take $N=L$ for half-filled case. 
In the atomic limit $t=0$, the phase boundary becomes a line $U=2V$ with $\Delta_{\rm c}=U(=2V)$. 
If finite $t$ is introduced, the system can gain some kinetic energy of the order of $t$ 
near the BOW-CDW instability due to the competition between the on-site and nearest-neighbor 
Coulomb interactions. Thus, the charge gap is minimized at the BOW-CDW boundary. 
Next, to evaluate the SDW-BOW boundary we calculate the spin gap
\begin{equation}
\Delta_{\rm s}=\lim_{L \to \infty}[E(N,1)-E(N,0)].
\end{equation}
If $V \ll U/2$, the system is a Mott insulator with $2k_{\rm F}$-SDW. The electrons are 
uniformly distributed over the system, so that there is no spin gap. As $V$ increases, 
the charge fluctuations are enhanced, and then a transition from the SDW phase to the 
BOW phase occurs. In the BOW phase, the electrons polarize alternatively and 
spin-singlet bound states are formed on dimers. Consequently, we can make an estimate 
of the SDW-BOW boundary as a point where the spin gap begins to develop. However, 
for some parameters the spin gap is too small to figure out if it remains finite, 
i.e., $\Delta_{\rm s} \lesssim 10^{-6}t$. Therefore, for verifying the presence of the spin gap 
we consider the spin-spin correlation function
\begin{equation}
S(q)=\frac{1}{L} \sum_{kl}e^{iq(k-l)}\left(\left\langle s^z_k s^z_l \right\rangle
-\left\langle s^z_k \right\rangle \left\langle s^z_l \right\rangle\right)
\end{equation}
with $q=2\pi/L$ and $s^z_i=n_{i\uparrow}-n_{i\downarrow}$. According to the Luttinger liquid 
theory~\cite{solyom79}, the long-range behavior of this function is governed by the LL spin 
exponents $K_\sigma$ ($=\lim_{q \to 0}\pi S(q)/q$). We find $K_\sigma=0$ in the spin-gapped phase 
and $K_\sigma=1$ everywhere else in the thermodynamic limit~\cite{Voit92}. This criterion enables 
us to estimate the SDW-BOW critical point precisely. Although we can obtain all the phase 
boundaries with the quantities mentioned above, the BOW oder parameter is also studied 
for making extra sure. The order parameter simply gives the boundaries between the BOW phase 
and the other phases. The BOW operator is given as
\begin{equation}
B_i = \frac{1}{2} \sum_\sigma 
(c^\dagger_{i \sigma}c_{i+1 \sigma} + c^\dagger_{i+1 \sigma}c_{i \sigma}).
\end{equation}
and we define the BOW order parameter $\langle B \rangle$ as an amplitude of the BOW 
oscillation in the center of the system, i.e., 
$\langle B \rangle=\lim_{L \to \infty}|\langle B_{L/2} - B_{L/2+1} \rangle|$. 
For $\langle B \rangle \neq 0$, a long-range order of the BOW state appears. 

\begin{figure}[htbp]
    \includegraphics[width= 7.8cm,clip]{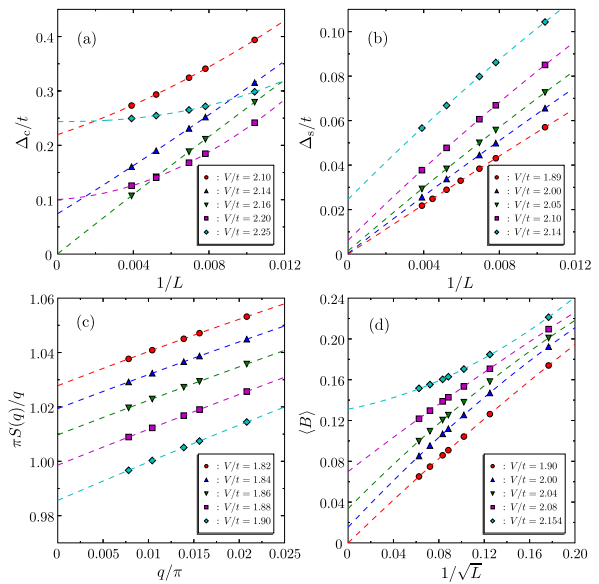}
  \caption{
Finite-size-scaling analyses for (a) the charge gap, (b) spin gap, 
(c) spin-spin correlation function, and (d) BOW order parameter near the phase boundaries 
at $U=4t$. 
  }
    \label{extrap}
\end{figure}

A careful extrapolation of these quantities is necessary to extract correct value in the 
thermodynamic limit $L \to \infty$. We thus study various lengths of chains with $L = 32$ to 
$512$ and perform finite-size-scaling analysis based on the $L$-dependence of the quantities. 
Figure \ref{extrap} shows the finite-size-scaling analyses for (a) the charge gap, (b) spin gap, 
(c) spin-spin correlation function, and (d) BOW parameter near the phase transitions at $U=4$. 
The charge (spin) gap is systematically extrapolated by performing a least-squares fit to the 
forth-order polynomial in $1/L$, reflecting the holon (spinon) band structure around 
the band edge. Then, an estimation of the LL spin exponent in the thermodynamic limit 
is not so simple for finite-size calculations. In the spin-gapless phase, one cannot expect 
easily find $K_\sigma \to 1$ exactly due to logarithmic corrections. However, the logarithmic 
corrections are known to vanish at which the spin gap opens, in analogy with the dimerization 
transition in the $J_1$$-$$J_2$ model~\cite{Egg96}. In the spin-gapped phase, there is 
a similar difficulty as follows; if the spin gap is small, the convergence 
of $K_\sigma$ to $0$ will obviously occur only for very large systems. As a result, we will 
estimate the critical point where the spin gap opens by $\pi S(q)/q$ crossing $1$ at $q \to 0$. 
This method was primarily used in Ref.~\cite{sengupta02}. Let us now turn to the BOW order 
parameter. Since the order parameter in the thermodynamic limit is very small compared to 
the finite-size results, a well-controlled finite-size extrapolation is mandatory. In our 
calculations, the most problematic finite-size effects are the Friedel oscillation due to the 
open edges. Assuming that the amplitude of the Friedel oscillation in the center of a finite 
chain scales as $L^{-K_\rho}$~\cite{white02}, the BOW order parameter would be well-extrapolated 
as a function of $1/L^{K_\rho}$. For example, we may expect $K_\rho \approx 0.5$ in the vicinity 
of the SDW phase, so that $\left\langle B \right\rangle$ is scaled better by $1/\sqrt{L}$ than 
by $1/L$ near the SDW phase.

\begin{figure}[htbp]
    \includegraphics[width= 6.8cm,clip]{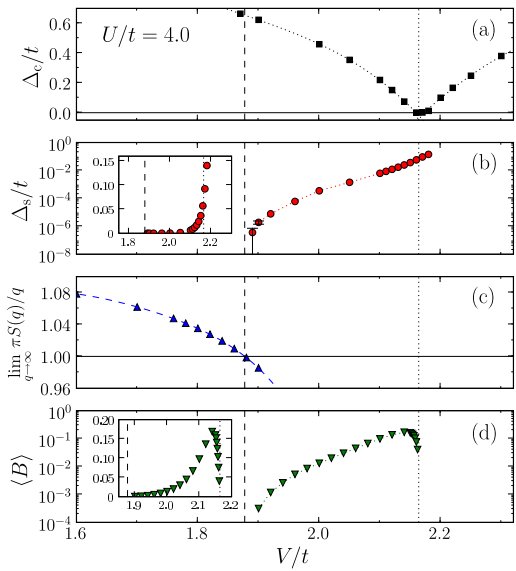}
  \caption{
Extrapolated results of (a) the charge gap, (b) spin gap, (c) spin-spin correlation function, and 
(d) BOW order parameter near the phase transition for $U=4t$. The dashed and dotted lines 
denote the SDW-BOW and BOW-CDW critical points, respectively. Insets: same quantities 
plotted with another scale.
  }
    \label{u4}
\end{figure}

Figure \ref{u4} shows the extrapolated results of (a) the charge gap, (b) spin gap, 
(c) spin correlation function, and (d) BOW order parameter around the phase transitions 
($U \sim 2V$) as a function of $V/t$ for $U=4t$. Let us look at the charge gap to 
estimate the BOW-CDW phase boundary. The charge gap decreases with approaching to 
a point $V \approx 2.164t$ and vanishes smoothly at the point. In other words, both the 
BOW and CDW insulating gaps start to develop gradually at the point. It means that 
a continuous transition between the BOW and CDW phases occurs at the critical point 
$V \approx 2.164t$. Note that the BOW insulating gap is of the nature of the Mott type.
We now turn to the SDW-BOW phase boundary. We find that the spin gap is finite for 
$V \gtrsim U/2$ and decreases with decreasing $V$. The critical point appears to 
lie around $V=1.9t$ from the disappearance point of the spin gap. The crossing point with 
$\pi \lim_{q \to 0}S(q)/q=1$ gives more precise estimation of the critical point 
$V \approx 1.877t$. Correspondently, the BOW order parameter has finite values only 
in the region $1.877t \le V \le 2.164t$. With increasing $V$, $\left\langle B \right\rangle$ 
rises exponentially from the SDW-BOW critical point, reaches the maximum value $\sim 0.18$ 
around $V = 2.14t$, and goes down to zero at the BOW-CDW critical point. Note that both 
values of the critical points are in good agreement with those of the previous quantum Monte 
Carlo (QMC) study~\cite{sandvik04}.

\begin{figure}[htbp]
    \includegraphics[width= 6.975cm,clip]{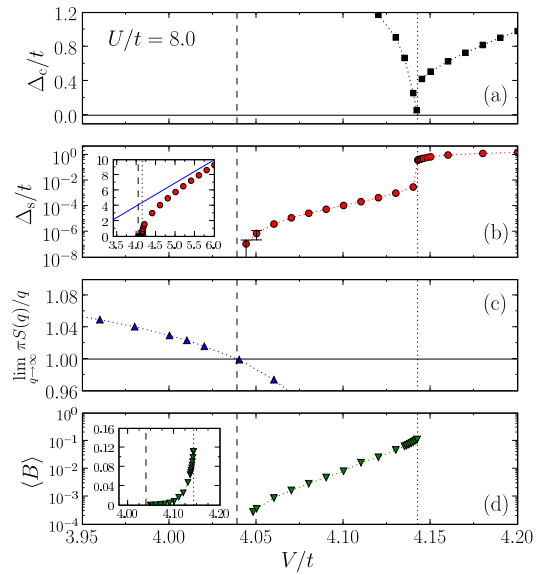}
  \caption{
The same quantities as in Fig.~\ref{u4} but for $U=8t$. Solid line in the inset of (b) 
denotes the spin gap in the $V/U \to \infty$ limit, i.e., $\Delta_{\rm s}=3V-U$.
  }
    \label{u8}
\end{figure}

Figure \ref{u8} shows the same quantities as in Fig.~\ref{u4} but for $U=8t$. 
Near the SDW-BOW phase boundary $V \approx 4.039t$, the behavior of all the quantities 
is qualitatively similar to those in the case of $U=4t$. On the other hand, the physical 
properties seem to be discontinuous at the BOW-CDW phase boundary $V \approx 4.142t$, 
which indicates that the transition is of first order. At the boundary, the charge gap 
remains finite and the slope of $\Delta_{\rm c}$ with respect to $V$ is discontinuous. 
However, the value of $\Delta_{\rm c}$ must be continuous since a competition between 
two kinds of charge configuration, i.e., CDW and uniform, leads to the BOW-CDW transition. 
Associated with this charge redistribution, the spin gap jumps by two orders of magnitude. 
In the CDW phase, it comes rapidly close to a line $\Delta_{\rm s}=3V-U$ which becomes exact 
in the $V/U \to \infty$ limit. Also, the BOW order parameter develops with approaching the 
BOW-CDW boundary and disappears at the transition point.

\begin{figure}[b]
    \includegraphics[width= 8.0cm,clip]{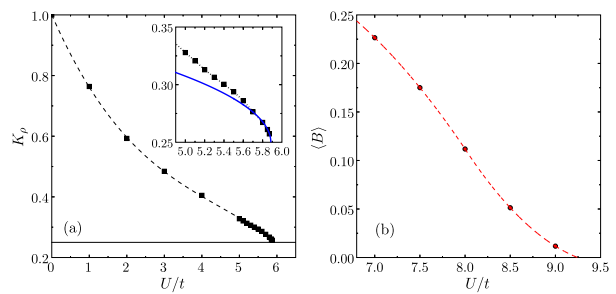}
  \caption{
Extrapolated results of the LL charge exponent (a) and the BOW order parameter (b) 
on the BOW-CDW boundary line. Inset: expanded view around the tricritical point 
$U_{\rm t}=5.89t$.
  }
    \label{critical}
\end{figure}

Whereas the BOW-CDW transition is continuous for $U=4t$, it is of first order for $U=8t$. 
Hence, a tricritical point $(U_{\rm t}, V_{\rm t})$, at which the transition changes from 
continuous to first order, must exist on the BOW-CDW boundary, as suggested in 
Refs.~\cite{eric02,sandvik04}. To evaluate the tricritical point, we examine the LL charge 
exponent $K_\rho$ via the derivative of charge structure factor at $q=0$~\cite{dzierzawa95}
\begin{equation}
K_\rho=\lim_{L \to \infty}\frac{1}{2}\sum_{kl}e^{i\frac{2\pi}{L}(k-l)}\left(\left\langle n_k n_l \right\rangle
-\left\langle n_k \right\rangle \left\langle n_l \right\rangle\right).
\end{equation}
Note that $K_\rho$ is finite only in the continuous Gaussian critical 
point~\cite{nakamura99,tsuchiizu02} for small $U$ and zero everywhere else. It was shown that 
the LL exponents can be obtained quite accurately with DMRG method~\cite{Eji05}. 
In Fig.~\ref{critical}(a), we plot DMRG results of $K_\rho$ as a function of $U/t$ on 
the BOW-CDW boundary line. As $U/t$ increases, $K_\rho$ decreases from $1$, reaches to 
$1/4$ at $(U_{\rm t},V_{\rm t})=(5.89t, 3.10t)$, and drops discontinuously to $0$; 
namely, a metal-insulator transition occurs at $U=U_{\rm t}$. Moreover, the $K_\rho$ curve is 
well-fitted by a function $K_\rho-1/4=0.061\sqrt{(U_{\rm t}-U)/t}$ near the tricritical point 
[see inset of Fig.~\ref{critical}(a)]. It implies that the transition is of the Kosterlitz-Thouless 
type. Let us now consider a point at which the BOW phase shrinks to $0$, which is called 
a ``critical end point''. The BOW state is still stable around the tricritical point and 
therefore the critical end point $(U_{\rm c}, V_{\rm c})$ would exist for $U_{\rm c}>U_{\rm t}$. 
For a fixed $U$ ($>U_{\rm t}$), the BOW order parameter has a maximum around the BOW-CDW 
boundary. To find the critical end point, we plot $\left\langle B \right\rangle$ on the BOW-CDW 
boundary as a function of $U/t$ in Fig.~\ref{critical}(b). $\left\langle B \right\rangle$ 
decreases with increasing $U/t$ and reaches to $0$ at $(U_{\rm c}, V_{\rm c})=(9.25t, 4.76t)$.
For $U \ge U_{\rm c}$, the transition is always first-order SDW-CDW one.

\begin{figure}[htbp]
    \includegraphics[width= 6.5cm,clip]{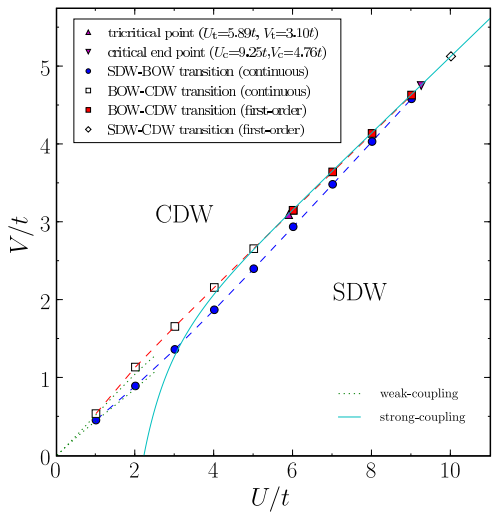}
  \caption{
DMRG phase diagram of the 1D half-filled EHM. The BOW phase exists between the SDW and 
CDW phases.
  }
    \label{pd}
\end{figure}

In Fig.~\ref{pd} we sum up our results as the ground-state phase diagram. One can see 
good agreement with the weak-coupling renormalization group (RG) results~\cite{tsuchiizu02} 
as well as the strong-coupling perturbation results~\cite{vandongen94}. The BOW phase has 
a maximum width at $U \sim 4t$, which is concerned with the fact that the effective 
nearest-neighbor exchange interaction is the largest at the intermediate couplings of $U$ in the 
half-filled Hubbard model~\cite{Szczech95}. It is so because the large exchange interaction 
promotes the formation of spin-singlet pair if the charge fluctuation is introduced by $V$. 
Accordingly, we confirm that the magnitude of the spin gap is maximized around $U \sim 4t$ 
in the BOW phase. 

In summary, we study the ground-state phase diagram of the 1D half-filled EHM using 
DMRG method. We calculate several quantities with considerable accuracy to determine 
the SDW-BOW and BOW-CDW boundaries. As for the phase boundaries, our data agrees 
quantitatively with the RG results in the weak-coupling regime ($U \lesssim 2t$), with 
the perturbation results in the strong-coupling regime ($U \gtrsim 6t$), and with 
the QMC results in the intermediate-coupling regime. We also find that the BOW-CDW 
transition changes from continuous to first order at the tricritical point 
$(U_{\rm t}, V_{\rm t})=(5.89t, 3.10t)$ and it locates far from the critical end point 
$(U_{\rm c}, V_{\rm c})=(9.25t, 4.76t)$. Since the previous DMRG results could be 
insufficient in accuracy, our results are not in agreement with them. We thus believe that 
our DMRG results bring a sound conclusion and put an end to the controversy on the phase 
diagram of the 1D half-filled EHM.

\acknowledgments

We thank R.M.~Noack, E.~Jeckelmann, and F.~Gebhard for useful discussions. We are grateful 
to M.~Tsuchiizu for his RG results and helpful discussions.


\begin{thebibliography}{99}
\bibitem{ishiguro90} T.~Ishiguro and K.~Yamaji, {\it Organic superconductors} 
(Springer-Verlag, Berlin, 1990).
\bibitem{kiess92} {\it Conjugated Conducting Polymers}, edited by H.~Kiess 
(Springer-Verlag, Berlin, 1992).
\bibitem{ishii03} H.~Ishii {\it et al.}, Nature (London) {\bf 426}, 540 (2003).

\bibitem{baeriswyl85} D.~Baeriswyl, in {\it Theoretical Aspects of Band Structures and 
Electronic Properties of Pseude-One-Dimensional Solids}, edited by R.\,H.~Kamimura 
(Reidel, Dordrecht, 1985), pp.~1-48.

\bibitem{emery79} V.\,J.~Emery, in {\it High Conducting One-Dimensional Solids}, 
edited by J.\,T.~Devreese, R.~Evrand, and V.~van Doren (Plenum, New York, 1979).
\bibitem{nakamura99} M.~Nakamura, J. Phys. Soc. Jpn. {\bf 68}, 3123 (1999); 
\prb {\bf 61}, 16377 (2000).

\bibitem{hirsch84} J.\,E.~Hirsch, \prl {\bf 53}, 2327 (1984).
\bibitem{cannon90} J.\,W.~Cannon and E.~Fradkin, \prb {\bf 41}, 9435 (1990); 
J.\,W.~Cannon {\it et al.}, \prb {\bf 44}, 5995 (1991).
\bibitem{vandongen94} P.\,G.\,J.~van Dongen, \prb {\bf 49}, 7904 (1994).
\bibitem{Voit95} J.~Voit, Rep. Prog. Phys. {\bf 58}, 977 (1995).
\bibitem{tsuchiizu02} M.~Tsuchiizu and A.~Furusaki, \prl {\bf 88}, 056402 (2002).
\bibitem{Tam06} K-M.~Tam {\it et al.}, \prl {\bf 96} 036408 (2006).

\bibitem{eric02} E.~Jeckelmann, \prl {\bf 89}, 236401 (2002).
\bibitem{sengupta02} P.~Sengupta {\it et al.}, \prb {\bf 65}, 155113 (2002).
\bibitem{sandvik04} A.\,W.~Sandvik {\it et al.}, \prl {\bf 92}, 236401 (2004).
\bibitem{zhang04} Y.\,Z.~Zhang, \prl {\bf 92}, 246404 (2004).
\bibitem{Glocke07} S.~Glocke {\it et al.}, preprint (cond-mat/0707.1015).

\bibitem{white92} S.\,R.~White, \prl {\bf 69}, 2863 (1992); \prb {\bf 48}, 10345 (1993).
\bibitem{solyom79} J.~S\'olyom, Adv. Phys. {\bf 28}, 201 (1979).
\bibitem{Voit92} J.~Voit, \prb {\bf 45}, 4027 (1992).
\bibitem{Egg96} S.~Eggert, \prb {\bf 54}, R9612 (1996).
\bibitem{white02} S.\,R.~White {\it et al.}, \prb {\bf 65}, 165122 (2002).
\bibitem{dzierzawa95} M.~Dzierzawa, {\it The Hubbard Model}, edited by D.~Baeriswyl {\it et al.}, 
NATO ASI Ser. B, Vol. {\bf 343} (Plenum Press, New York, 1995), pp.~327.
\bibitem{Eji05} S.~Ejima {\it et al.}, Europhys. Lett. {\bf 70}, 492 (2005).
\bibitem{Szczech95} Y.\,H.~Szczech {\it et al.}, \prl {\bf 74}, 2804 (1995).
\end{thebibliography}
\end{document}